\documentclass[aip,amsmath,amssymb,reprint,superscriptaddress,floatfix]{revtex4-1}

\usepackage{graphicx}% Include figure files
\usepackage{floatrow}
\usepackage[caption=false,font={large}]{subfig}
\usepackage{bm}% bold math

\usepackage[normalem]{ulem}
\usepackage{amsmath}
\newcommand{\stkout}[1]{\ifmmode\text{\sout{\ensuremath{#1}}}\else\sout{#1}\fi}

\usepackage{amssymb}
\usepackage{color}
\usepackage{times}
\usepackage{nicefrac}
\usepackage{float}
\usepackage[sort&compress]{natbib}

\usepackage{hyperref}

\hypersetup{
  colorlinks=true,
citecolor=black,
linkcolor=black,
urlcolor=black
  }

\usepackage[T1]{fontenc}
\usepackage[utf8]{inputenc}

\newcommand{\bdt}[1]{{\color{blue} #1}}
\renewcommand{\bdt}[1]{{\color{black} #1}}

\begin{document}

\title{Role of long jumps in L\'evy noise induced multimodality}

\author{Przemys{\l}aw Pogorzelec}
\email{pogorzelec.przemyslaw@gmail.com}
\affiliation{Doctoral School of Exact and Natural Sciences, Faculty of Physics, Astronomy and Applied Computer Science,
Jagiellonian University, \L{}ojasiewicza 11, 30-348 Krak\'ow, Poland}

\author{Bart{\l}omiej Dybiec}
\email{bartlomiej.dybiec@uj.edu.pl}

\affiliation{Institute of Theoretical Physics
and Mark Kac Center for Complex Systems Research,
Faculty of Physics, Astronomy and Applied Computer Science,
Jagiellonian University, \L{}ojasiewicza 11, 30-348 Krak\'ow, Poland}

%Lines break automatically or can be forced with \\

\date{\today}% It is always \today, today,
			 %  but any date may be explicitly specified

\begin{abstract}
L\'evy noise is a paradigmatic noise used to describe out-of-equilibrium systems.
Typically, properties of L\'evy noise driven systems are very different from their Gaussian white noise driven counterparts.
In particular, under action of L\'evy noise, stationary states in single-well, super-harmonic, potentials are no longer unimodal.
Typically, they are bimodal however for fine-tuned potentials the number of modes can be further increased.
The multimodality arises as a consequence of the competition between long displacements induced by the non-equilibrium stochastic driving and action of the deterministic force.
Here, we explore robustness of bimodality in the quartic potential under action of the L\'evy noise.
We explore various scenarios of bounding long jumps and assess their ability to weaken and destroy multimodality.
In general, we demonstrate that despite its robustness it is possible to destroy the bimodality, however it requires drastic reduction in the length of noise-induced jumps.
\end{abstract}

\pacs{02.70.Tt,
 05.10.Ln, %Monte Carlo methods statistical physics and nonlinear dynamics,
 05.40.Fb, % Random walks and Levy flights
 05.10.Gg, % Stochastic analysis methods (Fokker--Planck, Langevin, etc.)
  02.50.-r, % Probability theory, stochastic processes, and statistics
  }

%
%\vspace{2pc} \noindent{\it Keywords}: L\'evy flights, non-equilibrium stationary density, bimodality
\maketitle

\setlength{\tabcolsep}{0pt}

%%%%%%%%%%%%%%%%%%%%%%%%%%%%%%%%%%%%%%%%%%%%%%%%%%%%%%%%%%%%%%%%%%%%%%%%
%
%
\textbf{
Increasing number of observations demonstrates a plethora of situations in which non-Gaussian heavy tailed fluctuations are recorded.
This indicates the need to include heavy-tailed fluctuations in the description of dynamical systems especially in out-of-equilibrium realms.
The possibility of anomalously long jumps, drastically changes the properties of stochastic dynamical systems compared to their Gaussian noise driven counterparts.
It is the long jumps that lead to bimodal non-equilibrium stationary states in single-well, super-harmonic, potential wells.
Exploration of different scenarios for restricting long jumps, allows for a better understanding of how long jumps determine the properties of stationary states and how they affect their modality.
This provides a deeper understanding of the role of non-Gaussian long jumps and the mechanisms responsible for emergence of multimodal stationary states.
}

%%%%%%%%%%%%%%%%%%%%%%%%%%%%%%%%%%%%%%%%%%%%%%%%%%%%%%%%%%%%%%%%%%%%%%%%
%
%
\section{Introduction \label{sec:introduction}}

The motion in complex environments can be efficiently approximated by methods of stochastic dynamics \cite{schimanskygeier1997,klafter2012fractional,tome2016stochastic}, which incorporates stochastic processes into equations describing the system's dynamics.
The noise is used to efficiently and effectively approximate complex, not fully known interactions of the test particle with its environment.
The stochastic dynamics in a fixed potential is capable of producing bimodal stationary states.
Multimodal stationary states can emerge in multi-well potentials $V(x)$ under action of the additive Gaussian white noise (GWN), which is evident from the fact that stationary state \cite{gardiner2009,risken1996fokker,reichl1998} $p(x)$ in such a case is of Boltzmann--Gibbs type, i.e., $p(x) \sim \exp\left[-\beta V(x)\right]$.
This, somewhat obvious, mechanism corresponds to motion in an appropriately prepared medium in the equilibrium environment.
However, there are various mechanisms that can lead to emergence of multimodal stationary states which do not rely entirely on the shape of the potential.
The less intuitive, but more intriguing, approach involves utilizing different noise types to control the existence and modality of stationary states.
Such a setup allows one to obtain multimodal stationary states even in a single-well potential if the noise component is chosen appropriately.
For instance, L\'evy noise \cite{chechkin2002,chechkin2003, capala2019multimodal}, Ornstein--Uhlenbeck noise \cite{jacquet2018time} and fractional Gaussian noise \cite{guggenberger2021fractional} can lead to emergence of bimodal stationary states in single-well potentials.

Under action of Ornstein--Uhlenbeck and fractional Gaussian noises the multimodality is thought to be induced by the interplay of the correlation of the noise driven motion \cite{jacquet2018time,guggenberger2021fractional} and the action of the deterministic force.
\bdt{However, a very different mechanism of creating bimodality is observed for the L\'evy noise\cite{chechkin2006,chechkin2008introduction}, which produces long and independent jumps.}
The L\'evy noise is used to approximate  out-of-equilibrium systems \cite{dubkov2008,dybiec2009} displaying heavy-tailed fluctuations \cite{solomon1993,solomon1994,delcastillonegrete1998,shlesinger1993,klafter1996,chechkin2002b,delcastillonegrete2005,Katori1997,peng1993,segev2002,lomholt2005,Viswanathan1996,brockmann2006}.
It is capable of producing bimodal stationary states in super-harmonic, single-well potentials \cite{chechkin2002,chechkin2003}.
Clearly, in order to produce a bimodal stationary state in a single-well symmetric potential (under action of a symmetric noise), at the origin there has to be a minimum of stationary density.
Therefore, there must be a mechanism producing a deficit of the probability mass at the origin.
More precisely, under L\'evy driving, the multimodality is known to originate due to competition between two components of the system dynamics: (\textit{i}) long stochastic jumps and (\textit{ii}) deterministic sliding to the origin, which could be characterized by the diverging time.
Long jumps resulting in visits to distant points are produced by the heavy-tailed L\'evy noise, while sliding to the origin is governed by the deterministic force.
In a situation when a particle almost never returns to the origin before performing the next long jump the stationary density becomes bimodal.
Therefore, in the context of long jumps, a natural question arises: how long ``long jumps'' are needed in order to induce a multimodal stationary state in a single-well potential?
\bdt{The exploration of the role of long jumps is the main research question which we ask within the current manuscript.}

To explore the role of longs jumps on the formation of multimodal stationary states in a single-well potentials, we first enclose the system at hand in a box and investigate how the stationary states are affected by the box size and the way how the box edges interfere with the system dynamics.
This can be done in at least two ways: first by introducing reflecting boundaries (``reflection scheme''), second by rejecting jumps which would move the particle outside the box (``rejection scheme'').
Within the current study, we extensively compare both above mentioned options with respect to their role on the formation and robustness of bimodality.
Lastly, we provide comparison with the case of a truncated jump length distribution.
These scenarios significantly differ from reduction of spread of particles due to stochastic resetting \cite{evans2011diffusion,evans2020stochastic,gupta2022stochastic}, as (except for the ``truncation scheme'') they directly limit the accessible space which subsequently results in indirect modification of the jump length distribution.

\bdt{
Exploration of the role played by long jumps and possible scenarios of bounding them extends our understanding of anomalous dynamics in single-well potentials driven by L\'evy noises.
Action of non-Gaussian L\'evy noises can not only produce bimodal \cite{chechkin2002,chechkin2003,chechkin2004,dubkov2007} or multimodal \cite{capala2019multimodal} stationary states in super-harmonic single-well potentials  but is also capable of inducing multiple modes within sufficiently deep potential wells of multi-well potentials \cite{ciesla2019multimodal}.
Problem of modality of stationary states can be also studied in the higher number of dimensions \cite{szczepaniec2014stationary} or in the underdamped \cite{capala2019underdamped,capala2020nonlinear} regime.
In all those setups long jumps are responsible for emergence of multimodality.
Consequently, elimination or limitation of long jumps should affect properties of well known dynamical systems in the regime of anomalous-diffusion.
}

\bdt{
Processes driven by L\'evy noise, due to long pulses of random force, become discontinuous \cite{samorodnitsky1994,kilbas2006,podlubny1999}.
Discontinuity of trajectories is responsible for the difference between first passages and first arrivals, as jumps do not hit the target point but typically jump over it.
In turn it prevents application of well established methods of solving diffusion equation \cite{gardiner2009,risken1996fokker,chechkin2003b} and leads to the concept of leapovers  \cite{koren2007,koren2007b,wardak2020first}.
In general, it is not questioned that systems driven by L\'evy noise are described by the fractional variants of the diffusion equation\cite{jespersen1999,yanovsky2000,schertzer2001}.
However, the problems of properly defining\cite{kwasnicki2017ten,garbaczewski2019fractional}  (non-local) fractional operators \cite{kwasnicki2012eigenvalues,katzav2008spectrumfractional,zoia2007} or accounting for boundary conditions \cite{garbaczewski2019fractional,garbaczewski2022levy,wardak2020first} remain under active research.
}

The model under study is described in the next section (Sec.~\ref{sec:model} -- Models).
Obtained results are included in the Sec.~\ref{sec:results} -- Results and Discussion.
The paper is completed by Sec~\ref{sec:summary} -- Summary and Conclusions.
Additional information is moved to appendices:  \ref{sec:noise}  ($\alpha$-stable random variables and L\'evy noise)  and \ref{sec:sliding} (Deterministic sliding).

%%%%%%%%%%%%%%%%%%%%%%%%%%%%%%%%%%%%%%%%%%%%%%%%%%%%%%%%%%%%%%%%%%%%%%%%
%
%
\section{Models \label{sec:model}}
We study the general system described by the overdamped Langevin equation
\begin{equation}
    \frac{dx}{dt}=-V'(x)+\xi(t),
    \label{eq:langevin}
\end{equation}
where
\begin{equation}
    V(x)= \frac{1}{c} |x|^c\;\;\;\;\; (c>0)
    \label{eq:potential}
\end{equation}
is the single-well, power-law potential and $\xi(t)$ is a symmetric L\'evy ($\alpha$-stable) noise.
The $\alpha$-stable noise is a generalization of the Gaussian white noise to the non-equilibrium realms \cite{janicki1994}, where heavy tailed fluctuations are abundant \cite{ditlevsen1999b,mercadier2009levyflights,barkai2014,amor2016,barthelemy2008,fioriti2015,lera2018gross}.
The noise produces independent increments which follow a symmetric, unimodal, heavy-tailed $\alpha$-stable density \cite{janicki1994,samorodnitsky1994}, i.e., probability density with the characteristic function
$\varphi(k)  = \exp\left[ - \sigma^{\alpha}\vert k\vert^{\alpha} \right],$
where $\alpha$ ($0<\alpha \leqslant 2$) is the stability index controlling the asymptotics of the distribution, while $\sigma$ ($\sigma>0$) is the scale parameter \cite{janicki1994,samorodnitsky1994}.
For $\alpha=2$ one recovers the Gaussian white noise, while for $\alpha<2$ the non-Gaussian, heavy-tailed, power-law, L\'evy noise.
For more details see Refs.~\onlinecite{chechkin2002b,chechkin2006,dubkov2008} and App.~\ref{sec:noise}.
The steepness of the potential, see Eq.~(\ref{eq:potential}), is controlled by the exponent $c$.
\bdt{For $c>2$ the potential is super-harmonic, while for $0<c<2$ it is sub-harmonic.}

Under action of the L\'evy noise the evolution of the probability density $p(x,t|x_0,t_0)$ is governed by the fractional Smoluchowski--Fokker--Planck equation \cite{jespersen1999,yanovsky2000,schertzer2001}

\begin{equation}
\frac{\partial p}{\partial t} = \frac{\partial}{\partial x}\left[V'(x)p\right]+\sigma^\alpha \frac{\partial^{\alpha} p}{\partial |x|^{\alpha}}.
    \label{eq:fp-lf}
\end{equation}
The fractional Riesz--Weil $\frac{\partial^{\alpha} p}{\partial |x|^{\alpha}}$ derivative  \cite{podlubny1999,samko1993}  is understood in the sense of the Fourier transform, which is natural for unbounded domains
\begin{equation}
\mathcal{F}_k\left( \frac{\partial^\alpha f(x) }{\partial |x|^\alpha} \right)=-|k|^\alpha \mathcal{F}_k(f(x)).    
\label{eq:ft}
\end{equation}
However, in more complex setups, e.g., escape from a finite interval, other types of space-fractional derivatives are used \cite{podlubny1999,padash2019first} and additional care with respect to definitions of fractional operators and boundary conditions is required \cite{kwasnicki2017ten,song2017computing,cusimano2018discretizations}.
Subsequently, in the unbounded space and selected setups, from Eq.~(\ref{eq:fp-lf}), using the Fourier transform (characteristic function), the stationary state can be derived \cite{chechkin2003,chechkin2004}.

\begin{figure}
	\centering
    \includegraphics[width=0.98\columnwidth]{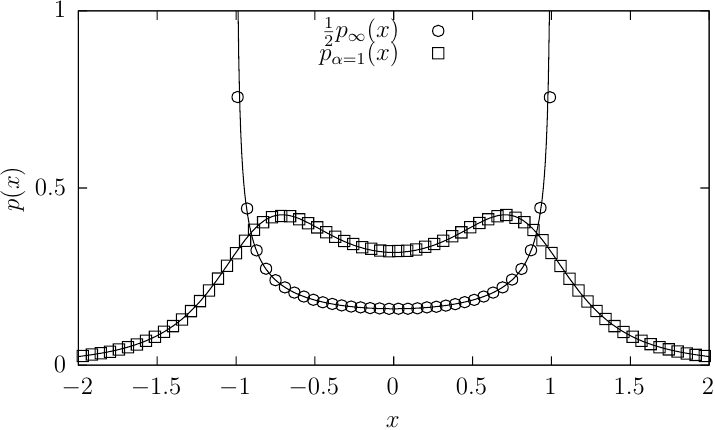}
	\caption{\bdt{Test of the Euler-Maruyama approximation, see Eq.~(\ref{eq:euler-levy}).
	Points represent results of numerical simulation of motion in the quartic-potential (empty squares) and the infinite rectangular potential well (empty circles), while solid lines depict theoretical stationary densities given by Eqs.~(\ref{eq:c4-stationary}) and (\ref{eq:rec-stationary}).
	For clarity of presentation $p_{\infty}(x)$ has been divided by the factor of 2.
	Numerical results have been averaged over $10^7$ realizations with $\Delta t=10^{-3}$ and $\varepsilon=10^{-3}$.}}
	\label{fig:em-test}
\end{figure}

Stationary states in systems driven by L\'evy noises exist for potential wells which are steep enough \cite{dybiec2010d}, i.e., $c > 2-\alpha$.
In even potentials $V(x)$, e.g., see Eq.~(\ref{eq:potential}), under action of symmetric noise they are symmetric.
For $c=2$ (harmonic potential) they are given by the $\alpha$-stable density, from which noise pulses were taken, with a changed (rescaled) scale parameter \cite{chechkin2002,chechkin2003,dybiec2007d}, which is the consequence of linearity of the Langevin equation in this case.
For $c>2$ stationary densities becomes bimodal \cite{chechkin2002,chechkin2003,dubkov2007,capala2019multimodal}.
In particular for $c=4$ and the Cauchy noise ($\alpha$-stable noise with $\alpha=1$), using the Fourier transform, see Eq.~(\ref{eq:ft}), it is possible to find the non-equilibrium stationary density\cite{chechkin2002,chechkin2003,chechkin2004,chechkin2006,chechkin2008introduction}
\begin{equation}
    p_{\alpha=1}(x)=\frac{1}{\pi\sigma^{\frac{1}{3}}\left[\left(\frac{x}{\sqrt[3]{\sigma}}\right)^4-\left(\frac{x}{\sqrt[3]{\sigma}}\right)^2+1\right]}
    \label{eq:c4-stationary}.
\end{equation}
\bdt{The stationary probability density $p_{\alpha=1}(x)$ given by Eq.~(\ref{eq:c4-stationary}) satisfies $\frac{d}{d x}\left[x^3p_{\alpha=1} (x)\right]+\sigma \frac{d }{d |x|} p_{\alpha=1}(x)=0$ equation, which is the Eq.~(\ref{eq:fp-lf}) after disregarding $\partial p /\partial t$ and setting $V(x)=x^4/4$ and $\alpha=1$.}
From Eq.~(\ref{eq:c4-stationary}) it is apparent that the stationary state is indeed a symmetric bimodal distribution, see Fig.~\ref{fig:em-test}, with modes at
\begin{equation}
x_{mod} =\pm \sigma^{\frac{1}{3}}/\sqrt{2}
\label{eq:sigma}    
\end{equation}
and the power-law asymptotics $p(|x|) \propto |x|^{-4}$.
The limit $c\to\infty$, see Eq.~(\ref{eq:potential}), transforms the single-well potential into the infinite rectangular potential-well with boundaries located at $x = \pm 1$.
In such a limit, the stationary density reads \cite{denisov2008}
\begin{equation}
p_{\infty}(x)=\frac{\Gamma(\alpha) (2L)^{1-\alpha}  (L^2-x^2)^{\alpha/2-1}}{\Gamma^2(\alpha/2)}
 \label{eq:rec-stationary}
\end{equation}
with $L=1$.
In contrast to the finite $c$, see Ref.~\onlinecite{dubkov2007}, the stationary density (\ref{eq:rec-stationary}) does not depend on the scale parameter $\sigma$.
Moreover, modal values are located at boundaries, i.e., at $\pm L$.

In order to assess the role of the long jumps on the emergence of bimodality we investigate three different ways of limiting jump lengths.
In scenarios ($i$) and ($ii$), the potential $V(x)$ is immersed in a $[-L,L]$ box, of variable half-width $L$, which introduces boundaries at $\pm L$, while in ($iii$) the jump length distribution is truncated at $\pm L$:

\begin{itemize}  
\item[($i$)] ``reflection scheme'': we introduce reflecting (impenetrable) boundaries at $\pm L$.
Too long jumps are shortened to reach at most $-L+\varepsilon$ or $L-\varepsilon$, see Ref.~\onlinecite{dybiec2017levy}, where $\varepsilon$ is a small ($\varepsilon \ll L$) positive parameter.
Consequently, the particle always stays within the $[-L,L]$ box.

\item[($ii$)] ``rejection case'':  we again introduce boundaries at $\pm L$, but now every jump which would move the particle outside the $[-L,L]$ box is rejected.
Such boundaries implement a space dependent truncation of the jump length distribution.
Note that this makes the jump length distribution space dependent and in general asymmetric as a particle can initiate a jump from any  $x \in (-L,L)$.
Furthermore, jump lengths are neither independent nor identically distributed.

\item[($iii$)]  ``truncation scheme'':  constraints are introduced by truncating the noise induced part of a jump at $\pm L$, that is jumps exceeding length $L$ are rejected.
Therefore, contrary to previous schemes, the particle can leave the $[-L,L]$ interval in a series of jumps (for jumps starting at $x=0$) or even in a single jump (for jumps starting at $x\neq 0$).
\end{itemize}
Note, that all the  considered schemes ultimately (directly or indirectly) affect the jump length distribution.
For scenario ($i$) and ($ii$) this is done indirectly by boundary conditions, while for ($iii$) the truncation is explicit.
\bdt{The possibilities presented and explored here do not exhaust the full spectrum of possible ways how to bound L\'evy flights by implementing various types of boundary conditions \cite{garbaczewski2019fractional} and how to treat underlying  processes on a single trajectory level\cite{dybiec2017levy,garbaczewski2022levy,wardak2020first}.
The ``truncation scheme'' implements truncated L\'evy flights\cite{mantegna1994b}, while rejection is an analog of taboo process \cite{garbaczewski2019fractional} and reflection is a censored like process\cite{garbaczewski2019fractional}.
The detailed discussion on the subtleties, similarities and differences among various possibilities of restricting L\'evy noise driven motion can be found in Ref.~\onlinecite{garbaczewski2019fractional}.
}

The Langevin equation (\ref{eq:langevin}) was integrated using the Euler-Maruyama method \cite{higham2001algorithmic,mannella2002}
\begin{equation}
    x(t+\Delta t) = x(t) -V'(x) \Delta t + \xi_t \Delta t^{1/\alpha},
    \label{eq:euler-levy}
\end{equation}
with the integration time step $\Delta t=10^{-3}$.
\bdt{
First of all, we have verified that for the uninterrupted motion the theoretical stationary states (the quartic potential with the Cauchy noise, see Eq.~(\ref{eq:c4-stationary}), and the infinite rectangular potential well, see Eq.~(\ref{eq:rec-stationary})) are correctly reconstructed, see Fig.~\ref{fig:em-test}.
Moreover, the integration time step $\Delta t$ was tested for self-consistency of the results, demonstrating that $\Delta t=10^{-3}$ is a reasonable choice.
}
The realizations of independent and identically distributed random variables $\xi_t$ following the symmetric $\alpha$-stable distributions have been generated using the GNU Scientific Library implementation of general methods \cite{chambers1976,weron1996} of generation of $\alpha$-stable random variables.
Nevertheless, studies have been limited to the Cauchy case ($\alpha$-stable density with $\alpha=1$) for which the general method \cite{chambers1976,weron1996}, reduces to the inversion of the cumulative density.
Additionally, we assume that the scale parameter $\sigma$ is set to $\sigma=1$.

\bdt{All bounding setups are explored by the Euler-Maruyama method, see Eq.~(\ref{eq:euler-levy}), however paths are affected by boundaries.
The ``reflection scheme'' accepts all jumps, but modifies the reached point, i.e., if the new positions is outside $[-L,L]$ the motion is stopped at $-L+\varepsilon$ (jump to the left) or $L-\varepsilon$ (jump to the right), see below.
For the ``rejection'' and ``truncation'' some jumps are rejected.
Scenarios ($ii$) and ($iii$) are realized by drawing jumps until the restraint condition is satisfied.
Such an approach corresponds to the so-called ``accept-reject'' method of generating (pseudo) random numbers \cite{devroye1986,barkema1999montecarlo}.
For ``rejection'' the jump is rejected if the new position would be outside the $[-L,L]$ box, while for ``truncation'' we do not accept jumps longer than $L$.
Numerical results have been averaged over $10^7$ (``truncation''), $2\times 10^7$ (``reflection'') and $10^8$ (``rejection'') trajectories.
From the ensemble of trajectories generated according to Eq.~(\ref{eq:euler-levy}),  accompanied by constraints imposed by studied scenarios, we have estimated time dependent densities $p(x,t)$.
After a sufficiently long time, a time dependent density approaches its stationary limit $p(x)$, i.e., the form which does not change in time (its time derivative vanishes).
For instance, the stationary solution given by Eq.~(\ref{eq:c4-stationary}) satisfies the fractional diffusion (\ref{eq:fp-lf}) equation after setting the time derivative $\partial p/\partial t$ to 0, see Ref.~\onlinecite{chechkin2003}.
Simulation time was adjusted in such a way that the stagnation of interquantile distance is observed, indicating that the stationary state has been reached.
Finally, in the ``reflection scheme'' $\varepsilon = 10^{-3}$ was used, as it is small enough to assure that in the infinite rectangular potential well the numerically constructed stationary state is given by Eq.~(\ref{eq:rec-stationary}), see Fig.~\ref{fig:em-test}.}

\bdt{
We conclude this section with some remarks on jump length distributions.
Each displacement consists of two parts: deterministic shift (induced by space dependent deterministic force) and random jump (produced by noise and  following the $\alpha$-stable distribution), see Eq.~(\ref{eq:euler-levy}).
Each jump is subject to additional constraints imposed by the assumed scenario of eliminating long leaps.
For the ``reflection'' and ``rejection'' jump length distribution is space dependent due to the fact that performed jumps are sensitive to the distance to boundaries.
For ``reflection'' we calculate a jump length, see Eq.~(\ref{eq:euler-levy}), which is effectively terminated when a particle crosses $-L$ or $L$ points.
The termination of jumps introduces the Dirac delta component to the jump length distribution, which is associated with stopping of particles at $\mp L  \pm \varepsilon$.
Heights of delta peaks are associated with the probability to overpass edges of the domain of motion to the left or right, i.e., they depend on the distances to boundaries.
For ``rejection'' the jump length distribution is restricted to the finite interval $[-L-x_0,L-x_0]$, where $x_0$ is a position before a jump.
Finally, for ``truncation'', random parts of jumps are independent and identically distributed.
They follow the $\alpha$-stable distribution truncated to $[-L,L]$.
Finally, the random part of successfully performed displacement is increased by the deterministic shift produced by the deterministic force.
}

%%%%%%%%%%%%%%%%%%%%%%%%%%%%%%%%%%%%%%%%%%%%%%%%%%%%%%%%%%%%%%%%%%%%%%%%
%
%
\section{Results and discussion\label{sec:results}}

Here, we present numerical results exploring properties of stationary states corresponding to three different schemes of restricting jump lengths described in Sec.~\ref{sec:model}.
In all cases, the magnitude of the restoring force $-V'(x)$ is implicitly reduced, because visits to points where the deterministic force is large are either limited or fully eliminated.
Moreover, the single-well potential $V(x)$, boundaries (scenario ($i$) and ($ii$)) and truncation (scenario ($iii$)) are symmetric.
Therefore obtained stationary states are symmetric as well, i.e., $p(x)=p(-x)$.
Following subsections correspond to: ``reflection scheme'' (Sec.~\ref{sec:reflection}), ``rejection scheme'' (Sec.~\ref{sec:rejection}) and ``truncation scheme'' (Sec.~\ref{sec:truncation}).

\begin{figure}[!ht]
\centering
\includegraphics[width=0.98\columnwidth]{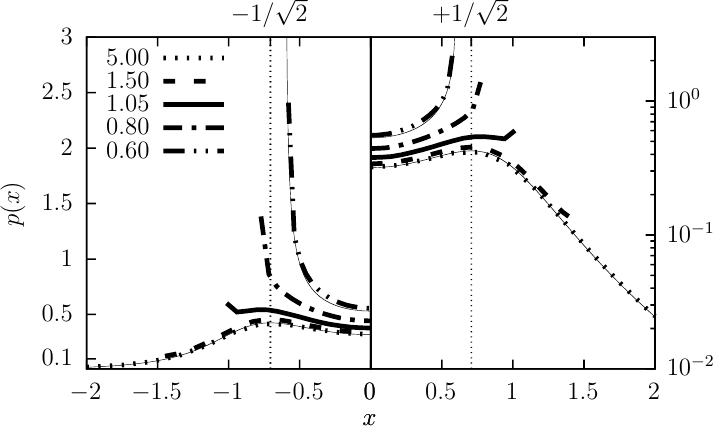}
\caption{Stationary probability densities for the quartic potential ($c=4$ in Eq.~(\ref{eq:potential})) with different values of $L$ under ``reflection scheme'' .
Jumps moving a particle to the exterior of $[-L,L]$ are terminated at $L-\varepsilon$ or $-L+\varepsilon$. $\varepsilon$ is set to $\varepsilon=10^{-3}$.
Left part of the plot is presented in linear scale, while in the right part the semi-log scale is used.
\bdt{Thin solid lines show stationary densities given by Eq.~(\ref{eq:c4-stationary}) with $\sigma=1$ and by Eq.~(\ref{eq:rec-stationary}) with $L=0.6$.}
}
\label{fig:cauchy-reflect-hists}
\end{figure}

%%%%%%%%%%%%%%%%%%%%%%%%%%%%%%%%%%%%%%%
%
\subsection{Reflection scheme\label{sec:reflection}}

\begin{figure}[!ht]
\centering
\includegraphics[width=0.98\columnwidth]{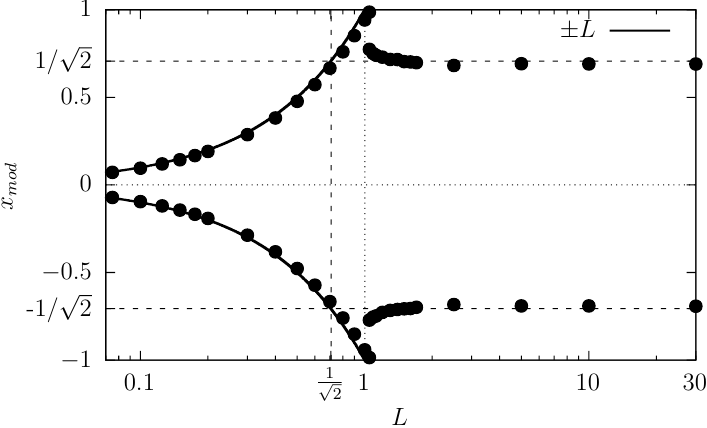}
\caption{The location of modes for the Cauchy noise and the quartic potential with different box half-widths $L$ under ``reflection scheme''.
\bdt{Solid lines show location of reflecting boundaries $\pm L$, which for small $L$ ($L<1$) agree with location of modes, i.e., $x_{mod} \approx \pm L$.
On the contrary, for large $L$ ($L>2$) modes are located at $\pm 1/\sqrt{2}$, which are indicated by dotted horizontal lines.
}
}
\label{fig:cauchy-reflect-modes}
\end{figure}

The ``reflection scheme'' implements impenetrable boundary conditions \cite{dybiec2017levy}.
 On the level of the Langevin equation,  this is accounted for by terminating jumps which would leave the $[-L,L]$ interval at $-L+\varepsilon$ or $L-\varepsilon$.
 As demonstrated in Refs.~\onlinecite{dybiec2017levy} and \onlinecite{kharcheva2016spectral}, for sufficiently small $\varepsilon$ ($\varepsilon \ll L$), such an approach properly implements impenetrable boundary conditions on the single trajectory level.
In particular, it correctly recovers \cite{dybiec2017levy} the stationary state in the infinite rectangular potential well, see Eq.~(\ref{eq:rec-stationary}) \bdt{and Fig.~\ref{fig:em-test}.}
\bdt{In Fig.~\ref{fig:em-test} we have used $\varepsilon=10^{-3} \ll L = 1$, which is sufficiently small to correctly reconstruct the theoretical stationary state.}
The infinite rectangular potential well can be also obtained as the $c\to\infty$ limit of the single-well potential $V(x)$, see Eq.~(\ref{eq:potential}).
 Consequently, the stationary state in the $c\to\infty$ limit is the same as the one given by Eq.~(\ref{eq:rec-stationary}), see Refs.~\onlinecite{dybiec2017levy,kharcheva2016spectral}.
 The stationary state in the infinite rectangular box  is always bimodal.
 The probability density monotonically grows with $|x|$  and reaches maxima at the boundaries.
For not too wide box, the deterministic force $-V'(x)$ in the $[-L,L]$  region is relatively small, and one could expect that once the box size $L$ becomes smaller than absolute value of modal value in the unrestricted space $x_{mod}$, i.e., $L<1/\sqrt{2} \approx 0.707$, modal values will be located approximately at $\pm L$.
However, it turns out that the situation is more intriguing.
We observe, see Figs.~\ref{fig:cauchy-reflect-hists} and~\ref{fig:cauchy-reflect-modes}, $x_{mod} \approx \pm L$ for all $L < 1$, i.e., modes are located at $\pm L$ already in situations when the box size is larger than the distance between modes for the uninterrupted motion.
As one can expect, a very similar situation is observed for $c=3$ or $c = 3.5$ (results not shown).
In contrast to the ``rejection scheme'' and ``truncation scheme'', which are studied below (Sec.~\ref{sec:rejection} and~\ref{sec:truncation}), the stationary states emerging  under the ``reflection scheme'' cannot be unimodal.
\bdt{For large L modes are located at $\pm 1/\sqrt{2}$ as for the uninterrupted motion, while for small $L$ they are at boundaries.}

Fig.~\ref{fig:cauchy-reflect-hists} shows stationary states for various box half-width $L$.
The model is symmetric with respect to $x\mapsto -x$ and the stationary states retain this symmetry.
Therefore, since various types of scales better illustrate different parts of the probability densities,
the left part of the Fig.~\ref{fig:cauchy-reflect-hists} is plotted in the linear scale, while the right part in the semi-log scale.
The same template is also used for presentation of stationary states under remaining schemes.
\bdt{For ``reflecting scheme'' stationary densities are restricted to $[-L,L]$ intervals.}

For $L \geqslant 2$ modes are located approximately \bdt{at the same position like for the uninterrupted motion, i.e.,   at $\pm 1/\sqrt{2}$} and the shape of stationary states are similar to the stationary density for the unrestricted (uninterrupted) motion.
This suggests that the impact of distant ``reflecting'' boundaries is negligible, because  accumulation of the probability mass in the vicinity of $\pm L$ with $L > 1$ is eliminated by the deterministic force.
For large $|x|$, the deterministic force $-V'(x)$ is significant and the probability mass is effectively moved towards the origin.
\bdt{In particular, for $L=5$, stationary density is practically given by Eq.~(\ref{eq:c4-stationary}) (which is depicted in Fig.~\ref{fig:em-test}) restricted to $[-L,L]$ interval, see superimposed dotted and thin solid lines in Fig.~\ref{fig:cauchy-reflect-hists}.}
The situation changes with the decreasing $L$, when the increasing accumulation of the probability mass at $\pm L$ starts to compete with the existing modes and modes are shifted towards the boundaries.
Interestingly, the $L\approx 1$ is a very special case as around $L \gtrapprox 1$ the outer modes transform to modes at $\pm L$ and the stationary is quadrimodal.
Once $L$ crosses $1$ the accumulation of the probability mass at the boundaries seems to overwhelm mechanisms producing the modes in the unrestricted space and the $x_{mod}\approx\pm L$ bimodal regime is reached.
For $L<1$ gluing of particles to the reflecting boundary is the main mechanism of modes creation, see solid lines in Fig.~\ref{fig:cauchy-reflect-modes} showing hypothetical dependence $x_{mod}=\pm L$.
Finally, for small $L$ the deterministic force is negligible and the action of reflecting boundaries is the only one mechanism producing bimodality, which is preserved for every $L$.
This is well visible for $L=0.6$ in Fig.~\ref{fig:cauchy-reflect-hists}.
\bdt{The recorded stationary density (dashed double-dotted line) is very close to the stationary state in the infinite rectangular potential well, see Eq.~(\ref{eq:rec-stationary}), depicted by the thin solid line, showing that the choice of $\varepsilon=10^{-3}$ was appropriate.}

\begin{figure}[!ht]
\centering
\includegraphics[width=0.98\columnwidth]{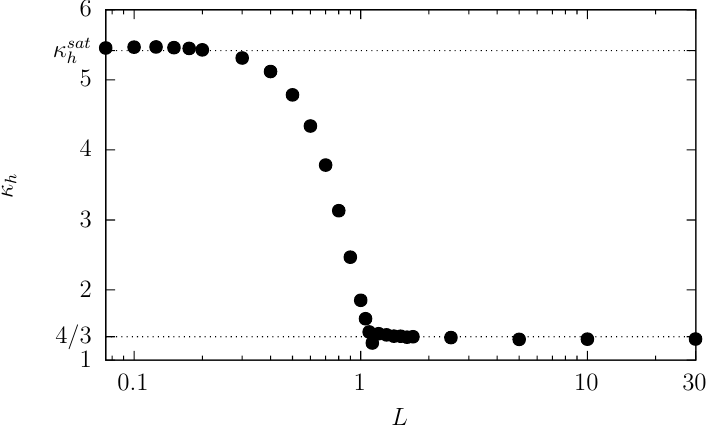}
\caption{
The ratio of heights $\kappa_h$: the quotient of height of lateral modes and the central minimum, see Eq.~(\ref{eq:ratioheight}), under the ``reflection scheme''.}
\label{fig:cauchy-reflect-ratio}
\end{figure}

The main case under study is the motion in a quartic potential.
Nevertheless, one can expect a similar behavior in other types of superharmonic potentials.
For instance, for the cubic potential $V(x)=|x|^3/3$ results (not shown) are qualitatively the same as for the quartic potential, see Fig.~\ref{fig:cauchy-reflect-modes}.
Quantitative differences are related to the fact that in the cubic potential modes are located closer to the origin than in the quartic potential.
\bdt{
Locations of modes are primarily determined by the deterministic force.
For the cubic potential, with $|x|<1$, the deterministic force is stronger than the  deterministic force produced by the quartic potential and modes emerge closer to the origin.
Consequently, the shift in the mode position produced by action of reflecting boundaries at $L \approx 1$ for $c=3$ is larger than for $c=4$ as probability mass is transferred more outward in comparison to the uninterrupted motion.}
\bdt{
In general, for smaller $c$ (constrained to the $c>2$, as for $2-\alpha<c\leqslant 2$ stationary densities are unimodal) modes are closer to the origin, while for increasing $c$ they move towards $\pm 1$ and in the $c\to\infty$ limit they are exactly at $\pm 1$.
The shift in modes’ locations is explained by the above-mentioned property of the deterministic force.}

In order to more deeply characterize the magnitude of bimodality of the stationary state one can calculate the ratio of heights $\kappa_h$, which is the quotient of average outer modes' height ($p(x_{mod})$) and the minimum of the stationary density ($p(0)$)
\begin{equation}
    \kappa_h = \frac{\frac{1}{2} \left[ p(-x_{mod}) + p(x_{mod}) \right]}{p(0)}.
	\label{eq:ratioheight}
\end{equation}
Fig.~\ref{fig:cauchy-reflect-ratio} shows the ratio $\kappa_h$ for the main studied case, i.e., for the quartic potential.
For large $L$ the ratio of heights saturates at
\begin{equation}
    \kappa_h = \frac{\frac{1}{2} \left[ p_{\alpha=1}\left(-\frac{1}{\sqrt{2}}\right)+p_{\alpha=1}\left(\frac{1}{\sqrt{2}}\right) \right]}{p_{\alpha=1}(0)} = \frac{4}{3},
\end{equation}
as for large box sizes the motion is practically uninterrupted, see Eq.~(\ref{eq:c4-stationary}).
With the decreasing $L$, $\kappa_h$ starts to grow and then saturates again.
For $L\approx 1$ a small drop can be seen, signaling emergence of $x_{mod}\approx \pm L$ modes of lower height than the inner ($x_{mod}\approx \pm \frac{1}{\sqrt{2}}$) ones.
For very small $L\to 0$, the motion approximately reduces to the free motion in an infinite narrow rectangular potential well.
Therefore, the $L\to 0$ limit of $\kappa_h$ can be calculated from Eq.~(\ref{eq:rec-stationary})
\begin{equation}
    \kappa_h = \frac{\frac{1}{2}\left[  p_{\infty}(-L) + p_{\infty}(L) \right]}{p_{\infty}(0)} = \lim_{x\to L} \frac{1}{\sqrt{1-\frac{x^2}{L^2}}}.
\end{equation}
Formally, the ratio of heights $\kappa_h$ diverges.
Nevertheless, as seen in Fig.~\ref{fig:cauchy-reflect-ratio}, $\kappa_h$ obtained from the simulation clearly saturates.
The numerically recorded saturation is caused by the discretization inherent to histograms.
In order to estimate the saturation level, we calculate:
\begin{eqnarray}
\langle\kappa_h\rangle & = &   \left\langle \frac{ p_\infty(x) }{p_\infty(0)}  \right\rangle_{[L-\Delta,L]}
 =    \frac{\langle p_\infty(x) \rangle_{[L-\Delta,L]}}{p_\infty(0)} \\ \nonumber
 & = & \frac{1}{\Delta}\int^L_{L-\Delta}\frac{dx}{\sqrt{1-\frac{x^2}{L^2}}},
\end{eqnarray}
where $\Delta$ is the bin width, i.e., $\Delta=2L/N$ with $N$ standing for the number of bins in the $[-L,L]$ interval.
After integrating one obtains
\begin{equation}
    \langle\kappa_h\rangle = \frac{N}{2}\left[\frac{\pi}{2}-\arcsin\left(1-\frac{2}{N}\right)\right].
    \label{eq:kappahestimate}
\end{equation}
The result depends on the number of bins, but, notably, does not depend on $L$.
As expected, Eq.~(\ref{eq:kappahestimate}), diverges with $\Delta\to 0$ ($N\to\infty$).
In our case $N=29$ and the estimated value $\langle \kappa_h \rangle \approx 5.42$ (dotted line in Fig. \ref{fig:cauchy-reflect-ratio}) is in good agreement with the results of simulations.
Qualitative dependence of $\kappa_h$ on the box half-width $L$ for the cubic (results not shown) and quartic potentials is the same.
Quantitative differences are due to differences in stationary densities.

%%%%%%%%%%%%%%%%%%%%%%%%%%%%%%%%%%%%%%%
%
% \clearpage
\subsection{Rejection scheme\label{sec:rejection}}

\begin{figure}[!ht]
\centering
\includegraphics[width=0.98\columnwidth]{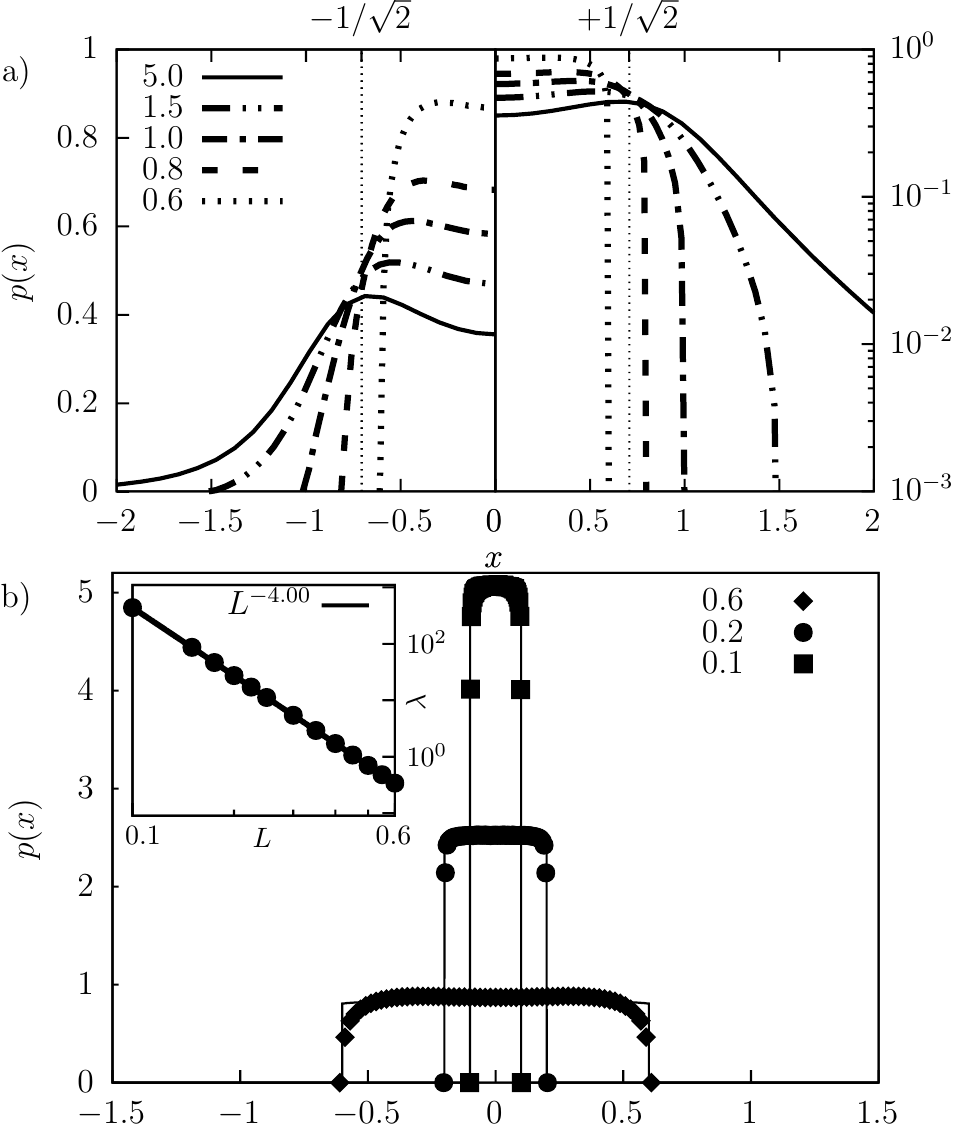}
\caption{Stationary probability densities for different values of $L$ under ``rejection scheme''.
In the top panel (a)), the left part of the plot is presented in linear scale, while in the right part the semi-log scale is used.
The bottom panel (b)) shows $\exp(-\lambda x^4)$ fits (solid lines) and values of the fitted exponent $\lambda$ (inset).
}
\label{fig:cauchy-reject-hists}
\end{figure}

\begin{figure}[!ht]
\centering
\includegraphics[width=0.98\columnwidth]{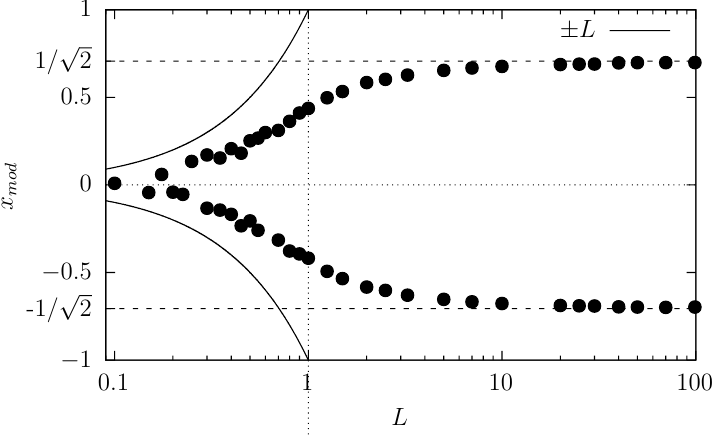}
\caption{The location of modes for the Cauchy noise for the quartic potential with different box sizes $[-L,L]$ under ``rejection scheme''.}
\label{fig:cauchy-reject-modes}
\end{figure}

\begin{figure}[!ht]
\centering
\includegraphics[width=0.98\columnwidth]{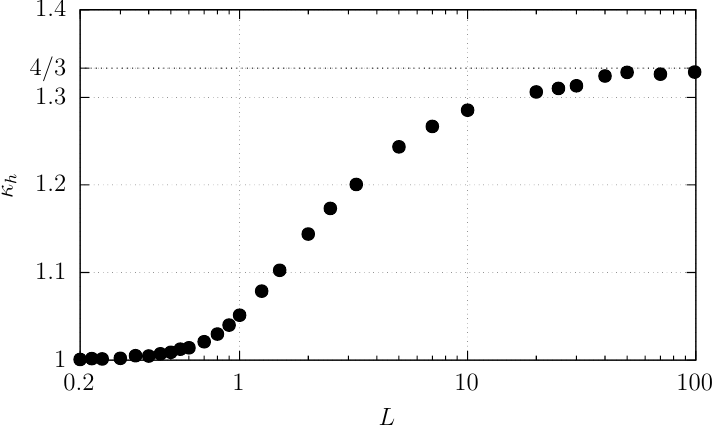}
\caption{
The ratio of heights $\kappa_h$, see Eq.~(\ref{eq:ratioheight}), under the ``rejection scheme''.}
\label{fig:cauchy-reject-ratio}
\end{figure}

The ``rejection scheme'' implicitly affects the jump length distribution and explicitly limits the accessible space to the $[-L,L]$ domain.
Analogously like for the ``reflection scheme'', we can expect that large enough boxes do not have the influence on modes positions, see Figs.~\ref{fig:cauchy-reject-hists} and~\ref{fig:cauchy-reject-modes}.
However, the critical box size is larger than for the ``reflection scheme''.
For $L \geqslant 5$ modes are located approximately at $\pm 1/\sqrt{2}$.
Decreasing of $L$ below $L=5$ noticeably, although continuously, shifts locations of modal values toward $0$, see Figs.~\ref{fig:cauchy-reject-hists} and~\ref{fig:cauchy-reject-modes} showing histograms and modes locations for $V(x)=x^4/4$.
Lastly, for sufficiently small $L$ the stationary density, under the ``rejection scheme'', becomes unimodal.
Fig.~\ref{fig:cauchy-reject-ratio} shows the ratio of heights $\kappa_h$, see Eq.~(\ref{eq:ratioheight}).
Analogously, like in Fig.~\ref{fig:cauchy-reflect-ratio} the ratio of heights saturates for large $L$ at  $\kappa_h=4/3$.
However, in contrast to the ``reflection scheme'', for small $L$, $\kappa_h$ tends to 1, because stationary densities with the decreasing $L$ first become uniform around the origin, i.e., outer modes become weaker, and finally become unimodal.

Similar to how rejecting of some jumps turns out to shift the modes towards the origin, see Fig.~\ref{fig:cauchy-reject-modes}, the same happens with simply reducing the strength of the noise in Eq.~(\ref{eq:langevin}).
The decay in $\sigma$ in the uninterrupted (without rejection) motion decreases the fraction of long jumps, although the power-law decay of the jump length distribution, see Eq.~(\ref{eq:asymptotics}) in App.~\ref{sec:noise}, is still characterized by the exponent $\alpha+1$.
Therefore, for the uninterrupted motion, with the decreasing $\sigma$ modes move linearly toward the origin, see Eq.~(\ref{eq:sigma}), and the stationary state always stays bimodal.

Under ``rejection scheme'' the bimodality arises for large enough $L$, because a domain of motion is sufficiently large that a particle can reach a point which is distant enough to the origin from which the time of deterministic return to origin is infinite, see App.~\ref{sec:sliding}.
Consequently, a particle cannot return to the vicinity of the origin before the next jump to a distant point.
For small $L$ the time of the deterministic return is still infinite but the diffusive spread compensates for it, resulting in emergence of the unimodal stationary state.
In other words, the diffusive spread is able to eradicate the minimum of the stationary density at the origin, which separates modes.
Unfortunately, due to fluctuations of the histograms with small $L$ (especially around $x\approx 0$) it is hard to indicate the exact value of the transition point.
Moreover, as demonstrated in Fig.~\ref{fig:cauchy-reject-hists}b) stationary densities are practically of the truncated Boltzmann--Gibbs type, i.e., $p(x) \propto \exp[-\lambda V(x)]$ restricted to the $[-L,L]$ interval.
The bottom part of Fig.~\ref{fig:cauchy-reject-hists} shows fitted densities (main plot) along with values of the fitted exponent $\lambda$ (inset) for the quartic potential.
The truncated Boltzmann--Gibbs like densities are recorded for very small $L$ since rejection allows for very short jumps only, which mimics independence and limits spatial dependence among jumps.
The fitted exponent $\lambda$ grows with the decrease in $L$, as for small $L$ stationary states are narrower.

The ``rejection scheme'' can be interpreted as a random truncation scheme, as box edges introduce truncation of long jumps.
Nevertheless, the level of truncation depends on the current position which determines the distance to the edge of the box and introduces the asymmetry to the effective jump length distribution.
The typical (sharp) truncation scheme is studied in the next subsection.

%%%%%%%%%%%%%%%%%%%%%%%%%%%%%%%%%%%%%%%
%
% \clearpage
\subsection{Truncation scheme\label{sec:truncation}}

\begin{figure}[!ht]
\centering
\includegraphics[width=0.98\columnwidth]{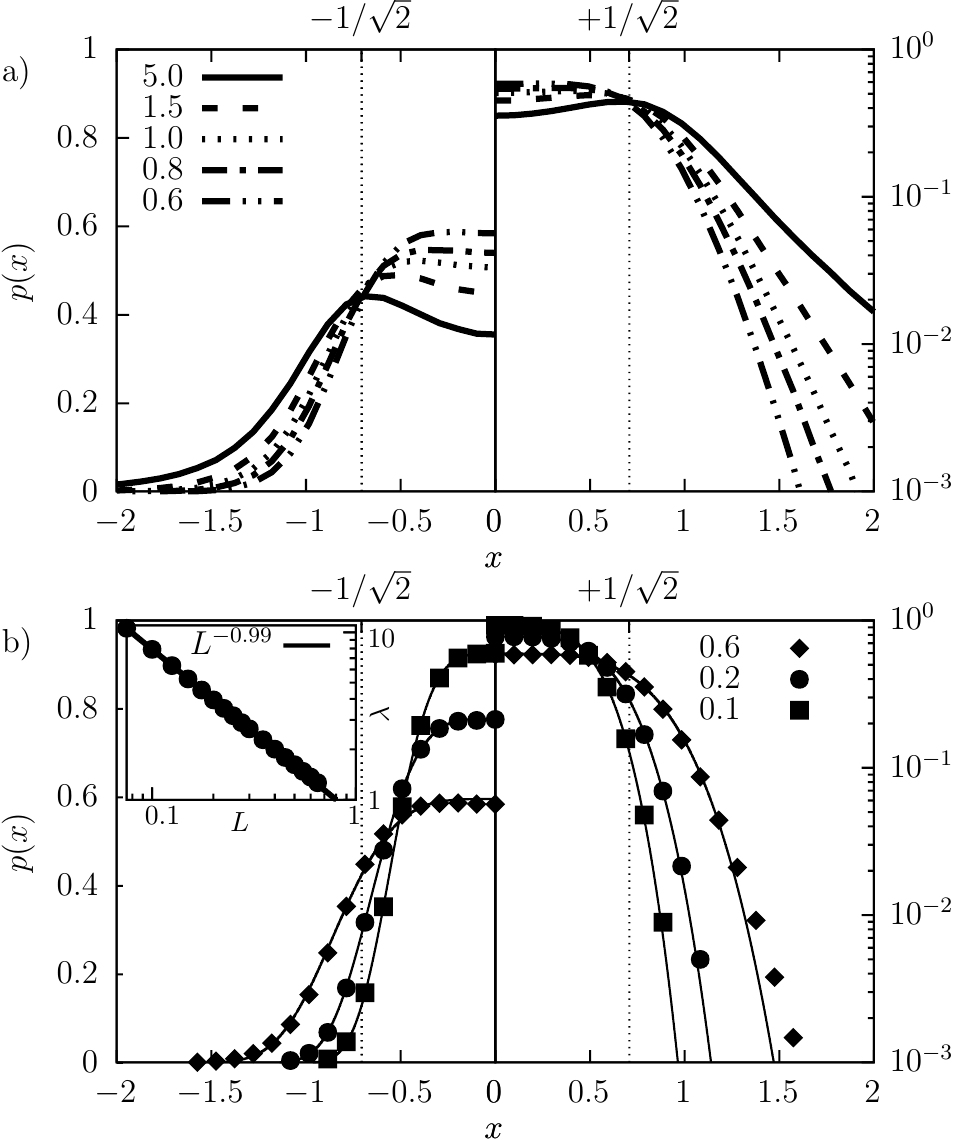}
\caption{Stationary probability densities for different values of $L$ under the ``truncation scheme''.
The bottom panel (b)) shows normalized $\exp(-\lambda x^4)$ curves fitted to numerically obtained histograms for  small values of $L$, with the inset presenting dependence of $\lambda$ on $L$.
In the top and bottom panels, the left part of the plot is presented in linear scale, while in the right part the semi-log scale is used.
}
\label{fig:cauchy-truncate-hists}
\end{figure}

\begin{figure}[!ht]
\centering
\includegraphics[width=0.98\columnwidth]{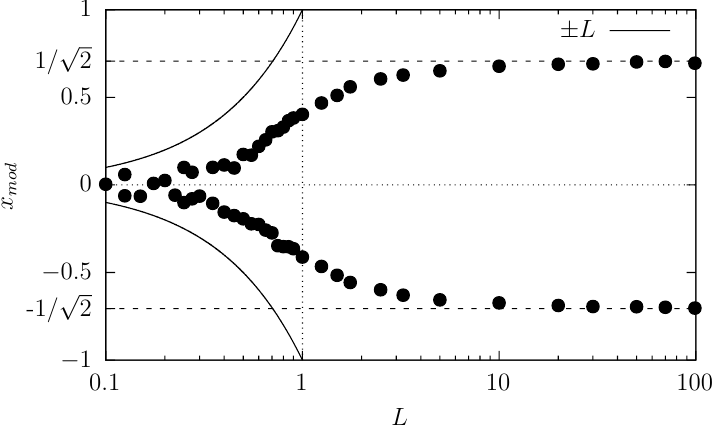}
\caption{The location of modes for the Cauchy noise and different box sizes $[-L,L]$ under the ``truncation scheme''.
}
\label{fig:cauchy-truncate-modes}
\end{figure}

\begin{figure}[!ht]
\centering
\includegraphics[width=0.98\columnwidth]{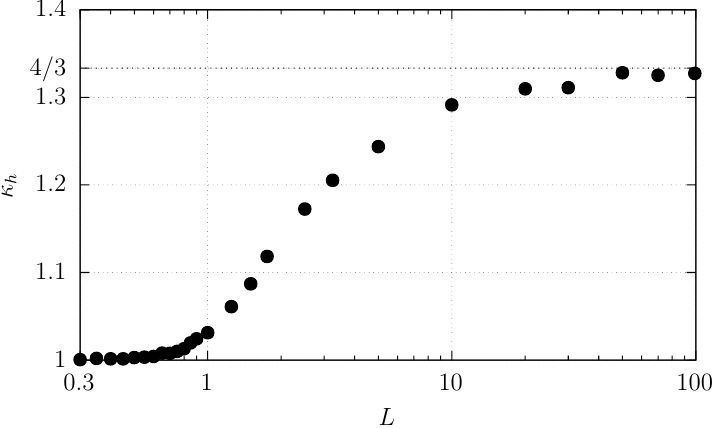}
\caption{
The ratio of heights $\kappa_h$, see Eq.~(\ref{eq:ratioheight}), under the ``truncation scheme''.
}
\label{fig:cauchy-truncate-ratio}
\end{figure}

In contrast to previously studied schemes, the ``truncation scheme'' does not limit the accessible space but directly cuts off the jump length distribution \cite{mantegna1994b}.
Only jumps  for which $|\xi_t \Delta t^{1/\alpha}| < L$ are accepted, see Eq.~(\ref{eq:euler-levy}).
Therefore, the $\alpha$-stable distribution of stochastic parts of jumps is truncated to the $[-L,L]$ interval.
For the ``\bdt{truncation} scheme'', constructed results are presented in a series of figures.
Fig.~\ref{fig:cauchy-truncate-hists} shows obtained histograms while Fig.~\ref{fig:cauchy-truncate-modes} location of modes.
Finally, Fig.~\ref{fig:cauchy-truncate-ratio}  displays the relative height of modes.

At the first glance the ``rejection'' and ``truncation'' schemes bear some similarities, as both of them affect the jump length distribution.
Nevertheless, they share fundamental differences which are mainly connected with the way how jumps are performed and how the accessible space is treated.
Only under the ``truncation scheme'' random components of jumps are independent and identically distributed.
In the ``rejection scheme'' a wandering particle cannot leave the $[-L,L]$ interval, while for the ``truncation'' it is possible.
Moreover, under the ``rejection scheme'', a particle located close to one of the boundaries can practically jump almost only in one (opposite) direction.
Nevertheless, for $L<1$, the rejection does not result in accumulation of the probability mass in the vicinity of boundaries.
In contrast, under the ``truncation scheme'', the  current position does not affect the jump length distribution.
The jump length is limited but the particle can reach a distant point from which it is efficiently returned to the neighborhood of the origin by the restoring force $-V'(x)$.
On the one hand, histograms for both setups are different, see Figs.~\ref{fig:cauchy-reject-hists} and~\ref{fig:cauchy-truncate-hists}.
In particular, they differ with respect to the domain (finite interval versus real line), which determines the large $|x|$ dependence and spread in the small $|x|$ range.
On the other hand, despite these fundamental differences, dependence of the modes location is qualitatively the same under both scenarios, compare Fig.~\ref{fig:cauchy-reject-modes} and Fig.~\ref{fig:cauchy-truncate-modes}.
Finally, Fig.~\ref{fig:cauchy-truncate-ratio} shows the ratio of  heights $\kappa_h$ as a function of $L$, which again is very similar to the dependence observed for the ``rejection scheme''.
For large $L$, the ratio saturates at $\kappa_h=4/3$, while in the $L\to 0 $ limits it tends to unity, because in the stationary states outer modes become weaker and finally disappear.

According to the central limit theorem \cite{feller1968,gnedenko1968} the sum of independent, identically distributed random variables characterized by a finite variance tends to the normal distribution.
Under the ``truncation scheme'', the $\alpha$-stable distribution of jumps is cut off, efficiently bounding noise induced jumps. 
Consequently, thanks to the central limit theorem, the acting noise can be approximated by the additive Gaussian white noise for which the emerging stationary density is of the Boltzmann--Gibbs type.
Indeed, for sufficiently small $L$, e.g., $L \leqslant 0.6$, the stationary density is approximately given by the Boltzmann--Gibbs distribution, i.e., $p(x) \propto \exp[-\lambda V(x)]=\exp[-\lambda x^4]$, see bottom panel of Fig.~\ref{fig:cauchy-truncate-hists}.
Additionally, the inset in Fig.~\ref{fig:cauchy-truncate-hists}b) displays fitted $\lambda$ as a function of the $L$.

\bdt{
The fact that the stationary density for small $L$ can be approximated by the Boltzmann-Gibbs distribution indicates that anomalously long jumps have been successfully eliminated and the acting noise resembles Gaussian white noise (with variance $\sigma_G^2$).  
For additive Gaussian white noise driving, the stationary density is of the Boltzmann-Gibbs type
$
p(x) \propto \exp[-V(x)/\sigma_G^2]
$
from which one gets  $\lambda \propto {\sigma_G^{-2}}$.
The prediction of $\lambda$ (or its scaling on the truncation level $L$) relies on estimation of the variance of jump length distribution, which due to elimination of long jumps becomes finite.
For the ``truncation'' case, the variance $\sigma^2$ of the (random part of) the jump length is given by the variance of the truncated L\'evy flights which scales as \cite{mantegna1994b,vinogradov2010cumulant}
\begin{eqnarray}
\sigma^2(\alpha,L) \propto  \frac{2\Gamma(1+\alpha) \sin\left(\frac{\pi\alpha}{2}\right)}{\pi(2-\alpha)}  L^{2-\alpha} .
\label{eq:tlf-variance}
\end{eqnarray}
Within the manuscript we use $\alpha=1$, therefore from Eq.~(\ref{eq:tlf-variance}) we get
\begin{equation}
\lambda \propto L^{-1},
\end{equation}
which is in accordance with the fit presented in the inset of Fig.~\ref{fig:cauchy-truncate-hists}b).
}

%%%%%%%%%%%%%%%%%%%%%%%%%%%%%%%%%%%%%%%%%%%%%%%%%%%%%%%%%%%%%%%%%%%%%%%%
%
%
\section{Summary and conclusions\label{sec:summary}}

The emergence of the multimodal stationary states under action of L\'evy noises is determined by the combined action of L\'evy flights and the deterministic force.
In steeper than parabolic potential wells, stationary states are always (at least) bimodal, because the time of deterministic return to the origin diverges.
Long excursions induced by L\'evy flights cannot be compensated by the restoring force and diffusive spread, which is produced by the central part of the jump length distribution.
Therefore, the stationary density attains the minimum at the origin.

In order to assess the role of long jumps in more detail we have studied several scenarios of limiting jump lengths.
The numerical studies have shown that elimination of long jumps or truncation of jump length distribution can destroy bimodality, however a significant cutback is required.
This indicates robustness of the bimodality in L\'evy noise driven systems.
\bdt{In other words, effects typically associated solely with L\'evy flights, which are characterized by the diverging variance, can be recorded in their truncated or censored versions, to name a few.
Such durability and generality increase chances of recording manifestation of long-jumps in various systems displaying heavy-tailed fluctuations.}
Introduction of reflecting boundaries removes exploration of distant points but is not capable of destroying bimodality of a stationary state due to the nature of impenetrable boundaries which supports emergence of modes at edges.
The accumulation of the probability mass at boundaries is especially strong for narrow intervals (centered at the origin) when the deterministic force  is negligible.

We have assumed that long excursions are reduced by disregarding long jumps without any time penalty.
On the operational level, jumps are drawn until one that satisfies the condition of constraints is drawn.
Such a scenario corresponds to the so-called ``accept-reject'' method of generating (pseudo) random numbers \cite{devroye1986}.
Importantly, the obtained results will stay intact for a finite average penalty/waiting time in analogous way like the continuous time random walk scenarios with finite mean waiting time produce the Markovian diffusion \cite{metzler2000} or finite return velocity does not  change the non-equilibrium stationary state under stochastic resetting \cite{evans2011diffusion,pal2019time}.
More general types of waiting times, i.e., power-law distributed, will not change the shapes of stationary states but affect the way how these states are reached in a similar manner like subdiffusion does not affect stationary states \cite{metzler2004} but only changes the way how they are approached.

%%%%%%%%%%%%%%%%%%%%%%%%%%%%%%%%%%%%%%%%%%%%%%%%%%%%%%%%%%%%%%%%%%%%%%%%
%
%
\section*{Acknowledgments}

We gratefully acknowledge Poland’s high-performance computing infrastructure PLGrid (HPC Centers: ACK Cyfronet AGH) for providing computer facilities and support within computational grants no. PLG/2023/016175 and PLG/2024/016969.
The research for this publication has been supported by a grant from the Priority Research Area DigiWorld under the Strategic Programme Excellence Initiative at Jagiellonian University.

\section*{Data availability}
The data (generated randomly using models presented in the paper) that supports the findings of this study is available from the corresponding author (PP) upon a reasonable request.

%%%%%%%%%%%%%%%%%%%%%%%%%%%%%%%%%%%%%%%%%%%%%%%%%%%%%%%%%%%%%%%%%%%%%%%%
%
%
\section*{Appendixes}
\appendix

The additional information presenting information on $\alpha$-stable random variables, L\'evy noise and deterministic sliding in a fixed potential is included in two appendices.
This information is useful for examination of the stochastic system and understanding the origin of bimodality.

%%%%%%%%%%%%%
\section{$\alpha$-stable random variables and L\'evy noise \label{sec:noise}}

The $\alpha$-stable noise is a generalization of the Gaussian white noise to the non-equilibrium realms \cite{janicki1994}.
The noise is still of the white type, i.e., it produces independent increments, but this time increments are distributed according to the heavy-tailed $\alpha$-stable density.
Within the current manuscript, we restrict ourselves to symmetric $\alpha$-stable noise only, which is the formal time derivative of symmetric $\alpha$-stable process $L(t)$, see Refs.~\onlinecite{janicki1994,dubkov2008}.
Increments $\Delta L=L(t+\Delta t)-L(t)$ of the $\alpha$-stable process are independent and identically distributed according to an $\alpha$-stable density.
Symmetric $\alpha$-stable densities are unimodal probability densities, with the characteristic function \cite{samorodnitsky1994,janicki1994}
\begin{equation}
\varphi(k)  = \exp\left[ - \sigma^{\alpha}\vert k\vert^{\alpha} \right].
    \label{eq:levycf}
\end{equation}
The stability index $\alpha$ ($0<\alpha \leqslant 2$) determines the tail of the distribution, which for $\alpha<2$ is of the power-law type
\begin{equation}
p(x) \propto |x|^{-(\alpha+1)}.
\label{eq:asymptotics}
\end{equation}
The scale parameter $\sigma$ ($\sigma>0$) controls the width of the distribution, typically defined by an interquantile width or by fractional moments $\langle |x|^\kappa \rangle$ ($0<\kappa<\alpha$), because the variance of $\alpha$-stable variables diverges \cite{samorodnitsky1994,weron1995} for $\alpha<2$.

The characteristic function of the symmetric $\alpha$-stable density, see Eq.~(\ref{eq:levycf}), reduces to the characteristic function of the normal distribution $N(0,2\sigma^2)$ for $\alpha=2$,
\begin{equation}
    f_2(x)=\frac{1}{\sqrt{4\pi \sigma^2}} \exp\left[ -\frac{x^2}{4\sigma^2} \right] ,
\end{equation}
while for $\alpha=1$ one gets the Cauchy distribution \cite{janicki1994}
\begin{equation}
    f_1(x)= \frac{\sigma}{\pi} \frac{1}{\sigma^2+x^2}.
\end{equation}
In other cases, $\alpha$-stable densities can be expressed using special functions \cite{penson2010,gorska2011}.

%%%%%%%%%%%%%
\section{Deterministic sliding\label{sec:sliding}}

\noindent
The Langevin equation
\begin{equation}
    \dot{x}(t) = -V'(x) + \xi(t)
\end{equation}
in the zero noise limit for $V(x)=|x|^{c}/c$ ($c>0$) reduces to
\begin{equation}
    \dot{x}(t) = -\mathrm{sign}(x) |x|^{c-1}
    \label{eq:newton}.
\end{equation}
From Eq.~(\ref{eq:newton}) one can find the dependence of $x(t)$ ($c \neq 2$)
\begin{equation}
    x(t)= \left[ x_0^{2-c} + (c-2) t \right]^{\frac{1}{2-c}},    
\end{equation}
while for $c=2$

\begin{equation}
    x(t)=x_0 \mathrm{e}^{-t}.
\end{equation}
From $x(t)$ one can deduce that for any $x_0$ with $0<c<2$ the time of deterministic sliding to the origin is finite ($t=x_0^{2-c}/(2-c)$), while for $c \geqslant 2$ it diverges, because $\lim\limits_{t\to\infty}x(t)=0$.
Therefore, the deterministic motion in power-law potential changes its properties at $c=2$.
Moreover, the transition observed for $c > 2$ plays an important role in noise driven systems as it is one of components required to induce bimodality of stationary states in single-well potentials.
More precisely, if the diffusive spread is unable to compensate for the diverging return time the stationary density becomes bimodal.

% \bibliography{apssamp}% Produces the bibliography via BibTeX.

\section*{References}
% \bibliography{core-bibliography}

%merlin.mbs apsrev4-1.bst 2010-07-25 4.21a (PWD, AO, DPC) hacked
%Control: key (0)
%Control: author (72) initials jnrlst
%Control: editor formatted (1) identically to author
%Control: production of article title (-1) disabled
%Control: page (0) single
%Control: year (1) truncated
%Control: production of eprint (0) enabled
\def\url#1{}

\end{document}